\documentstyle[prc,twocolumn,aps,epsf]{revtex}       

\begin{document}
\draft

\twocolumn[%
\newlength{\abswidth}
\setlength{\abswidth}{\linewidth}
\addtolength{\abswidth}{-.6in}
\begin{center}
\begin{minipage}{\abswidth}
{\large\bf{Spectroscopy of Double-Beta and Inverse-Beta Decays from $^{100}$Mo
for Neutrinos \\}}

\begin{center}
H.~Ejiri$^1$, 
J.~Engel$^2$, 
R.~Hazama$^1$, 
P.~Krastev$^3$, 
N.~Kudomi$^4$, 
and 
R.G.H.~Robertson$^1$ \\

{\it 
$^1$Nuclear Physics Laboratory and Dept. Physics, University of Washington, Seattle, WA 98195, USA\\
$^2$Dept. Physics and Astronomy, University of North Carolina,NC 27599, USA\\
$^3$Dept. Physics, University of Wisconsin, WI 53706,USA\\
$^4$Research Center for Nuclear Physics, Osaka University, Ibaraki, 
Osaka 567-0047, Japan\\
}
\end{center}

\hspace{3mm}
 Spectroscopic studies of two $\beta$-rays from $^{100}$Mo are shown to be of potential 
interest for investigating both the Majorana $\nu$ mass by neutrinoless double
$\beta$-decay(0$\nu\beta\beta$) and  low energy solar $\nu$'s by inverse
$\beta$-decay. With a multi-ton $^{100}$Mo detector, coincidence studies of
correlated $\beta\beta$ from 0$\nu\beta\beta$, together with the large $Q$
value($Q_{\beta\beta}$), permit identification of the $\nu$-mass term with a
sensitivity of $\sim$ 0.03 eV.  Correlation studies of the inverse
$\beta$ and the successive $\beta$-decay of $^{100}$Tc, together with the large
capture rates for  low energy solar $\nu$'s, make it possible to detect in
realtime individual low energy solar $\nu$ in the same detector.
\hspace{2cm}
\end{minipage}
\end{center}]

   Neutrino mass is a key issue of current neutrino($\nu$) physics.
Recent results with atmospheric \cite{SK}\cite{IMB}, solar
\cite{SK}\cite{GALLEX}, and  accelerator \cite{LSND} neutrinos strongly suggest
$\nu$ oscillations due to non-zero $\nu$-mass differences  and flavour mixings.
Neutrino oscillation measurements, however, do not give the  $\nu$ masses
themselves. The minimum $\nu$ mass consistent with the accelerator-$\nu$
oscillation is in the  eV range \cite{LSND}. The minimum mass associated with
the atmospheric-$\nu$ effect is of the order of 0.05 eV \cite{SK}. Neutrino
mass of astroparticle interest is in the range of 1$\sim$0.01 eV \cite{mina}.
It is of great interest to study directly $\nu$ mass with sensitivity down to 
$\sim$0.03 eV. 
  
   Double beta decay may be the only probe presently able to access such  small
$\nu$ masses.  Actually, observation of neutrinoless double beta decay
(0$\nu\beta\beta$) would identify a Majorana-type electron $\nu$ with a
non-zero {\em effective} mass $<m_{\nu}>$ \cite{doi}-\cite{tomoda} . 
  Calorimetric measurements of total $\beta\beta$-energy spectra have been made
on $^{76}$Ge , $^{130}$Te and other isotopes  \cite{doi}\cite{bau}\cite{ale}.
They give upper limits on 
$<m_{\nu}>$ in the sub-eV to eV region. 
  The 0$\nu\beta\beta$ process is, in fact, sensitive not  only to the $\nu$ mass 
($<m_{\nu}>$) but also to a right-handed weak current and other terms beyond
the Standard Model(SM)
\cite{doi}\cite{ejiri}\cite{faess}.  Spectroscopic studies of the energy and
angular correlations for two $\beta$-rays are useful to identify the terms
responsible for 0$\nu\beta\beta$. Spectroscopic measurements for two
$\beta$-rays have been made on $^{82}$Se, $^{100}$Mo, $^{136}$Xe and on others
\cite{doi}\cite{sre}-\cite{leu}. They give upper limits of a few eV on
$<m_{\nu}>$.  NEMO III will study $<m_{\nu}>$ in the sub-eV region \cite{nemo}.
Backgrounds(BG) from radioisotope (RI) impurities make it hard to perform
spectroscopic studies with  sensitivities down to $\sim $0.05 eV. 

    Solar neutrinos have been  studied  for more than 30 years \cite{bahpin}.
Low-energy solar-$\nu$ studies, so far, have been carried out with $^{71}$Ga
and $^{37}$Cl detectors \cite{GALLEX}. They are non-realtime and inclusive
measurements that do not identify the $\nu$ sources in the sun. Realtime
spectroscopic studies of low energy solar $\nu$ are important for studies of
the solar-$\nu$ problems
\cite{bahpin}-\cite{ragha}. They require, however, extremely low RI
impurities of the order of $b \sim 10^{-5}$ Bq/ton or less 
\cite{borex}.  Delayed coincidence studies with $\gamma$-rays are an excellent
way to reduce BG as proposed by Raghavan\cite{in}\cite{ragha}.  

   Rates for both the 0$\nu \beta \beta$ decay at $<m_\nu > \sim $0.03 eV and  the
inverse-$\beta$ decay induced by solar $\nu$ are extremely small, 6$\sim 8$
orders of magnitude smaller than BG rates for normal  two neutrino $\beta
\beta(2\nu
\beta \beta)$ and RI impurities  with $\sim $Bq/ton. Nuclei used for $\beta 
\beta$ decays have a potential for solar $\nu$ studies
\cite{ragha}. It is interesting and important  to find nuclei with adequate
0$\nu \beta \beta$ and solar-$\nu$-capture rates,  and effective ways to select
the rare 0$\nu \beta \beta$ and  inverse $\beta$ signals   from much larger BG
signals due to 2$\nu \beta \beta$ and  RI impurities. 

 The present Letter  shows that it is possible by measuring two correlated
$\beta$ rays from  $^{100}$Mo to perform both spectroscopic studies of
0$\nu\beta\beta$ with a sensitivity of the order of 
$<m_{\nu}>\sim$0.03 eV, and realtime  exclusive studies of low energy solar
$\nu$ by inverse $\beta$ decay. The unique features are as follows.

\begin{figure}[hbt]
\vspace*{-0.8cm}
\epsfxsize=8cm \epsfysize=6.5cm \epsfbox{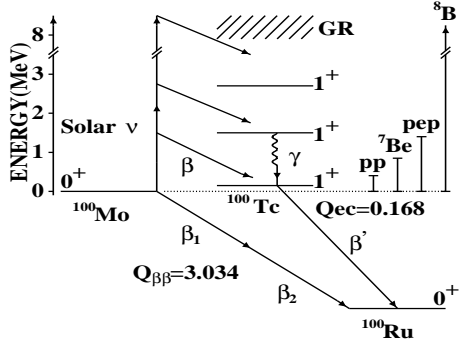}
\vspace*{-0.8cm}
\caption[]
{Level and transition schemes of $^{100}$Mo for double beta decays  ($\beta_1\beta_2$) and two beta decays ($\beta\beta^{'}$) induced by 
solar-$\nu$ absorption. GR is the Gamow-Teller giant resonance. $Q_{\beta\beta}$ 
and $Q_{\rm{ec}}$ are given in units of MeV.} 
\label{decy}
\end{figure}

1)The $\beta_1$ and $\beta_2$ with the large energy sum of $E_1+E_2$ are
measured  in coincidence for the 0$\nu\beta\beta$ studies, while the inverse
$\beta$-decay  induced by the solar
$\nu$ and the successive $\beta$-decay are measured sequentially in an adequate
time window for the low energy solar-$\nu$ studies. The isotope $^{100}$Mo is just
the one that satisfies the conditions for the $\beta\beta-\nu$ and
solar-$\nu$ studies, as shown in Fig.~\ref{decy}. 
 
2)The large $Q$ value of $Q_{\beta\beta}$=3.034 MeV gives a large phase-space
factor $G^{0\nu}$  to enhance the 0$\nu\beta\beta$ rate  and a large energy sum
of $E_1 + E_2 = Q_{\beta\beta}$  to place the 0$\nu\beta\beta$ energy signal
well above most BG except $^{208}$Tl and
$^{214}$Bi. The energy and angular correlations for the two $\beta$-rays can be
used to  identify the $\nu$-mass term. 
 
3)The low threshold energy of 0.168 MeV for the solar-$\nu$ absorption allows
observation of low energy sources such as pp and $^7$Be. The GT strength to 
the $1^+$ ground state of $^{100}$Tc is measured to be 
$(g_A/g_V)^2B(GT)$=0.52$\pm$0.06  by both charge-exchange reaction and 
electron capture \cite{aki}\cite{ec}. Capture rates are large even for low
energy solar
$\nu$'s, as shown in Table 1. The rates are 1.1 and 3.3 per day for $^7$Be $\nu$
and pp $\nu$, respectively, for 10 tons of $^{100}$Mo. The solar-$\nu$ sources
are identified by measuring the inverse-$\beta$ energies.  Only the
$^{100}$Tc ground state can absorb
$^7$Be $\nu$ and pp $\nu$. 

 4)The measurement of two $\beta$-rays (charged particles) enables one to
localize in space and in time the decay-vertex points for both the
0$\nu\beta\beta$ and solar-$\nu$ studies.
Radiations associated with BG are also measured. The
tightly localized $\beta$-$\beta$ event in  space and  time  windows, together
with relevant
$\beta$ and $\gamma$  measurements, are key points for selecting
$0\nu\beta\beta$ and solar-$\nu$ signals and for reducing  correlated and
accidental BG by factors $10^{-5}\sim 10^{-6}$ \cite{kudo}.\\  

  The 0$\nu\beta\beta$ transition rate $R_{0\nu}$ for $<m_{\nu}>$ is given by 
\begin{equation} 
R_{0\nu} = G^{0\nu} ( M^{0\nu} )^2 |<m_{\nu}>|^2,
\end{equation} 
where $G^{0\nu}$ is the phase space factor and $M^{0\nu}$ is the matrix element
\cite{doi}
\cite{ejiri}\cite{faess}, both relatively large for $^{100}$Mo.
 
\begin{table}
\caption[]{
Solar-$\nu$ absorption rates $R_{\nu}$ for $^{100}$Mo.} 
\vspace{.3cm}
\begin{tabular}{cccc}
Source  & $E_{\nu}^{(max)}$(MeV) & $E_{\beta}^{(max)}$(MeV) & $R_{\nu}$/SNU$^{a}$ \\ \hline
  pp  &   0.42   &     0.25    &           639 $\pm$ 85\\
  pep &   1.44   &     1.27    &            13 $\pm$ 2\\
  $^7$Be &   0.86 &    0.69    &           206 $\pm$ 35 \\
  $^8$B  &  $\sim$15  &   $\sim $14.2   &           27(23)$^{b} \pm$ 4 \\
  $^{13}$N &   1.20  &   1.03  &            22$\pm$ 3 \\
  $^{15}$O  &  1.74  &   1.57  &            32  $\pm$ 4\\ 
\end{tabular}
 $E_{\nu}^{(max)}$, $E_{\beta}^{(max)}$ are the maximum $\nu$ energy and maximum $\beta$-ray energy, respectively. \\
a) Standard-solar-model(SSM) capture rates based on BP98\cite{bahpin} with errors from those of $B(GT)$. 
b) Rate for the states below the effective neutron threshold energy. 
\end{table}
 
The $g^\nu_{7/2} - g^\pi_{9/2}$ shell-model structure of $^{100}$Mo -
$^{100}$Tc leads to the large measured 2$\nu\beta\beta$ rate \cite{doi,kudo}, and large
calculated values for the 0$\nu\beta\beta$ transition rate
\cite{faess,tomoda}.

  The 0$\nu\beta\beta$ events are identified by setting the appropriate energy
window and the prompt time window for the  $\beta\beta$ coincidence signals.
The rate in units of 10$^{-36}$/s  is
given as $ R_{0\nu}$=6.6$\times 10^4|<m>|^2$/(eV)$^2$  by  RQRPA \cite{faess}.
 The uncertainty in calculation of  the matrix element is considered to be 
 of order 50 \% \cite{tomoda}.

  For solar $\nu$ detection, the inverse $\beta$-decay  induced by the solar-$\nu$ 
absorption is followed by  $\beta$-decay with a mean life $\tau$ = 23 sec. 
Thus a time window can be set as
$\Delta T$=30 sec(10$^{-6}y$) from $t_1$ = 1 sec to $t_2$ = 31 sec. The
starting time of 1 sec is long enough to reject most  correlated BG such as the
2$\nu\beta\beta$, $\beta$-rays followed by conversion electrons, scatterings of
single $\beta$-rays, etc. The stopping time of  31 sec is  short enough to
limit  the accidental coincidence BG. The  accidental rate is further
reduced by effectively subdividing the detector into  $K$  unit
cells by means of position readout.

   Signal and background rates for 0$\nu \beta \beta$ and for $^7$Be and pp solar- $\nu$'s
are summarized  in Table \ref{tab2}. There are 8 efficiencies implicitly in Table \ref{tab2}. The efficiency
$\epsilon_{0\nu}$ arises from the 0$\nu \beta \beta$ window(cut) optimized for the
$\nu$ mass term in 0$\nu \beta \beta$, and is approximately 0.14. Here $\beta _1$
and $\beta _2$ are required to be oppositely directed with energies larger than 0.5 MeV.
The efficiency $\epsilon_{2\nu}$ describes the degree to which
$2\nu$ events fall in the $0\nu$  window, and is found to be $1.2 \times 10^{-8}$ by Monte Carlo
calculations \cite{kudo} with an assumed energy resolution
 $\Delta E/E \sim 7\%$  in FWHM at $E_1 + E_2 =Q_{\beta \beta }=$ 3.034  MeV.  The efficiency
$\epsilon_{\rm U} \simeq 0.6 \times 10^{-5}$ reflects the $\beta $ branch of $^{214}$Bi
in the 0$\nu \beta \beta$ window without preceeding decays from $^{214}$Pb being detected.
The efficiency for  $^{208}$Tl  with the effective $Q_{\beta}\sim$1 MeV is negligible.
It is noted that signals of $\gamma$ rays following the $^{208}$Tl $\beta$ decay
are separated in space from the $\beta$ signal, and thus are rejected, and the
internal conversion coefficients are very small. 

\begin{table}[tb]
\caption[]{
Expected rates for signals and backgrounds per ton of $^{100}$Mo isotope(9.6\% of natural Mo) 
in a natural Mo detector containing 15\% Mo by weight. Rates for U and Th are disintegration rates 
for $^{238}$U and $^{232}$Th} 
\vspace{.3cm}
\begin{tabular}{cccc}
Source & Raw Rate & Effective & Effective \\
 & /ton $^{100}$Mo/y & ($0\nu\beta\beta$) & (solar $\nu$) \\
\hline
$0\nu$, 0.05 eV & 31  &  4.3  &  - \\
$2\nu$, $1.15~10^{19}$ y & 3.6$\times$10$^{8}$ &  $\sim 4.4$ & - \\
10$^{-13}$ U w/w & 2.6 $\times$ 10$^6$ & $\sim$ 15 & - \\
10$^{-13}$ Th w/w & 8 $\times$ 10$^5$ & $\sim 0$  & - \\
                 &             &           &   \\
$^7$Be (SSM) & 39 & - & $ \sim 14$ \\
$pp$ (SSM) & 121 & - & $\sim 26$ \\
Correlated & 2.6 $\times$ 10$^6$ & $\sim 0$ & $\sim$7$\times$ 10$^{7} \Delta T $  \\
Accidental & 3.6 $\times$ 10$^{8}$ & $\sim$ 0 & $\sim$1.6 $\times$ 10$^{15} \Delta T/K $ 
\label{tab2}
\end{tabular}
10$^{-13}$(0.1ppt) $^{238}$U$\sim 1.25~10^{-3} b$(Bq/ton) for $^{214}$Bi.\\ 
10$^{-13}$(0.1ppt) $^{232}$Th$\sim 0.45~10^{-3} b$(Bq/ton) for $^{208}$Tl.
\end{table}

 The two solar neutrino efficiencies,
$\epsilon_{7} = 0.35$  and $\epsilon_{pp} = 0.2 $, are dominated by losses in
the Mo foils. The correlated BG comes mainly from the
successive $\beta$-decays of $^{214}$Pb$\rightarrow ^{214}$Bi
$\rightarrow ^{214}$Po for which an efficiency $\epsilon_c \simeq 27\times \Delta T$ is
estimated from the ground-state branch of $^{214}$Bi. The efficiency $\epsilon_a$ for 
the acidental BG, which is mainly due to the 2$\nu \beta \beta $, is estimated as $\epsilon_a
\sim 4.3\times 10^{6}\Delta T/K$ with $1/K$ being the spatial resolution. 

  The lower limit (sensitivity) on  $<m_{\nu}>$ can be obtained
by requiring that the number of  0$\nu\beta\beta$ events  has to exceed the
statistical fluctuation of the BG events.  The  sensitivity of the order of
$<m_{\nu}> \sim$ 0.03 eV can be achieved for three year measurement by means of a 
realistic  detector with a few tons of $^{100}$Mo and  RI  contents of the order of
0.1ppt($b\sim 10^{-3}$Bq/ton), in contrast to $b \sim 10^{-6}$Bq/ton  for calorimetric methods. 
 Sensitivity for the solar $\nu$ 
is obtained similarly as in case of the 0$\nu \beta \beta$. It is of the order
of $ \sim 100$ SNU for one year measurement by using the same   detector with  $K\sim
10^9$. In fact the $2\nu \beta \beta$ rate and the BG rate from RI at 0.1ppt( $b\sim 10^{-3}$Bq/ton)  
are larger than the solar $\nu$ rate by  factors $\sim 10^7 $ and $\sim 10^5$, respectively.
The fine localization in time($\Delta T=10^{-6}y$) and in space(1/$K=10^{-9}$),
which is possible with the present two-$\beta$ spectroscopy, is crucial for
reducing BG rates in realistic detectors. 
 
   The cosmogenic isotopes to be considered are long lived $^{93}$Mo, and short
lived $^{99}$Nb and $^{100}$Nb. Although $^{93}$Mo isotopes are not removed 
chemically,  they decay by emitting very low energy X-rays and conversion
electrons. Their energies can be lower than the detector threshold. $^{99}$Nb
and $^{100}$Nb are produced by fast neutrons at underground laboratories and
decay within  tens of  seconds by emitting
$\beta$-rays. They are estimated to give negligible contributions to the
present energy and time windows. The $^{14}$C and other single $\beta$ BG's
become negligible in the present two $\beta$ detection.

  One possible detector is outlined below. It uses a supermodule of 3.3
tons of $^{100}$Mo (34 tons of Mo) purified to $10^{-3}$ Bq/ton
for $^{238}$U and $^{232}$Th or less. This purity level has been achieved for
Ni and other materials for the Sudbury Neutrino Observatory\cite{ncd}. An
ensemble of plastic scintillator modules is newly designed on the basis of
recent developments \cite{min,kon,kudo}. The  supermodule with a  fiducial volume of
$(x,y,z)$=(6m,6m,5m) is composed of 1950  modules with $(x,y,z)$ =
(6m,6m,0.25cm). Each module may consist of 30 extruded plastic bars of
(x,y,z)=(6m,0.2m,0.25cm).

  The Mo foils with thickness  of 0.05 g/cm$^2$ are interleaved between the
modules. Light outputs from each scintillator module are collected by 222
WLS(wave length shifter) fibers with 2.7 cm interval for the $x$ direction at
the front side of the plane and the same for $y$ at the back side, each with
1.2mm in diameter and 6 m in length. An extrapolation of the results of
\cite{min,kon} suggests the large fraction of the WLS with respect to the
scintillators and the arrangement in both $x$ and
$y$ directions may give adequate photo-efficiencies and  energy resolutions. The
attenuation along the WLS can be corrected for by reading the position.  The
total 8.66  10$^5$ WLSs are viewed through clear fibers by 6800 16-anode
PMTs (photomultiplier tubes) with a large photo-electron efficiency, each
accepting 128 fibers with 8-fold multiplexing. All PMTs are well
separated through the clear fibers from the scintillator to avoid BG from
PMTs.

\begin{figure}[hbt]
\epsfxsize=7.5cm \epsfysize=4.8cm \epsfbox{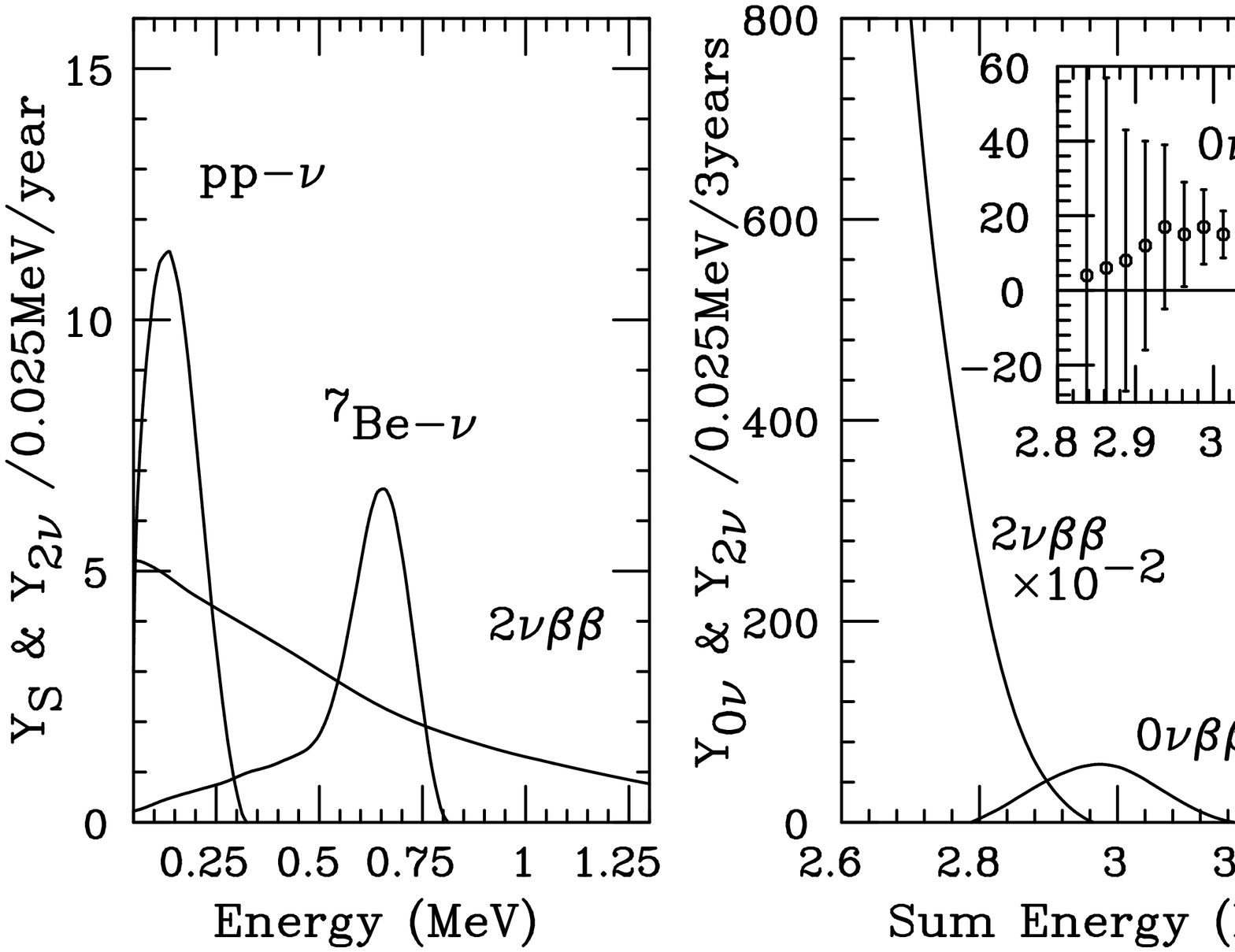}
\ \vspace*{0.1cm}
\caption[]
{Schematic energy spectra for a possible detector  with 3.3 tons of $^{100}$Mo (see text).
Right hand side: Sum energy spectra for 3 year measurement of 2$\nu\beta\beta$ and 0$\nu\beta\beta$ for $<m_{\nu}> = 0.1$ eV with $M^{0\nu}$ in \cite{faess}. The inset shows the 0$\nu\beta\beta$ spectrum  with $<m_{\nu}>$=0.05 eV after correction for 
2$\nu\beta\beta$ with statistical errors.
Left hand side: Inverse $\beta$ spectra for $^7$Be $\nu$ and pp $\nu$ on the basis of SSM\cite{bahpin}, and possible 2$\nu\beta\beta$ backgrounds. The detector threshold is set at 50 keV}
\label{kudo-haza}
\end{figure}


 The  scintillator ensemble designed gives a time resolution of
$\Delta  T/T \sim$ 1 ns in FWHM, and adequate energy and spatial  resolutions of
$\Delta E/E\sim 0.125 /\sqrt E(MeV)$ in FWHM and  $\Delta x=\Delta y\sim 0.5$
cm /$\sqrt E(MeV)$ with $\pm 2\sigma$. Then the energy resolutions of 7, 15, and 
30 $\%$, and the spatial ones of 1/$K$=0.11$\times 10^{-9}, 0.55 \times 10^{-9}$ and 
2.5$\times 10^{-9}$ are expected for 0$\nu \beta \beta $, $^7$Be-$\nu$ and pp-$\nu$, respectively.
  Expected energy spectra   are shown schematically for 0$\nu\beta\beta$ and solar $\nu$,
together with the major remaining BG of $2\nu
\beta \beta$ in Fig.~\ref{kudo-haza}.

 The detector can be used also for supernova $\nu$ studies and other rare
nuclear processes, and for other isotopes. Another option is
a liquid scintillator \cite{orl} in place of the solid one, keeping similar
configurations of the WLS readout. The energy and spatial resolution are
nearly the same. Then $^{150}$Nd with the large $Q_{\beta \beta}$  may be used either
in  solid  or  solution in the liquid scintillator for 0$\nu \beta \beta$.
Of particular interest is $^{136}$Xe because liquid Xe is a scintillator.

 The present spectroscopic  method for 0$\nu \beta \beta$ measures selectively
the $\nu $ mass term. In view of the dependence of $R_{0\nu}$ on $M^{0\nu}$
 calculations, 0$\nu \beta \beta$ studies on different nuclei are important. 
Thus the present method is complementary to the calorimetric $\beta
\beta$ measurements with high energy resolutions for
$^{76}$Ge \cite{bau} and $^{130}$Te \cite{ale}. The present method for solar $\nu$
gives adequate yields for  both pp-$\nu_e$ and $^7$Be-$\nu_e$ via charged current 
interaction, and their ratio is independent of the $ B(GT)$ value.
BOREXINO \cite{borex}, which requires much higher purity of $b\sim10^{-6}$ Bq/ton,
will give large yields for $^7$Be-$\nu_x$ and higher energy $\nu_x$ with $x=e,\mu ,\tau$
mainly via the charged current interaction and partly via the neutral current interaction.
LENS \cite{ragha} is 
sensitive to $^7$Be-$\nu$ and pp-$\nu$ charged currents with different technique and 
relative yields. Thus the present method provides important data for 
solar $\nu$,which are supplementary to 
existing geochemical and planned realtime experiments.  

The authors thank the Nuclear Physics Laboratory and Institute for Nuclear
Theory, University of Washington for support and discussions, and  Profs.
A.~Konaka, R.~Raghavan, and P.~Vogel for valuable comments and encouragements. R.~H
is supported by a Japanese Society for Science Promotion Fellowship.

\end{document}